\documentclass[amsfonts, amssymb, amsmath, preprint, showkeys, nofootinbib]{revtex4-1}
\usepackage[english]{babel}
\usepackage[utf8]{inputenc}
\usepackage[colorinlistoftodos, color=green!40, prependcaption]{todonotes}
\usepackage{amsthm}
\usepackage{mathtools}
\usepackage{physics}
\usepackage{xcolor}
\usepackage{graphicx}
\usepackage[left=23mm,right=13mm,top=35mm,columnsep=15pt]{geometry} 
\usepackage{adjustbox}
\usepackage{placeins}
\usepackage[T1]{fontenc}
\usepackage{lipsum}
\usepackage{csquotes}
\usepackage{verbatim}
\usepackage[pdftex, pdftitle={Article}, pdfauthor={Author}]{hyperref} 

\def\be{\begin{equation}}
\def\ee{\end{equation}}
\def\ber{\begin{eqnarray}}
\def\eer{\end{eqnarray}}
\def\bwt{\begin{widetext}}
\def\ewt{\end{widetext}}

\def\cH{{\cal H}}
\def\f{\frac}

\begin{document}

\title{Quantum field theory in a de-Sitter universe transiting to the radiation stage}
\author{Juan R. Salazar}
\email{jsalazar16@ucol.mx}
\affiliation{Facultad de Ciencias - CUICBAS, Universidad de Colima, Colima, C.P. 28045, M\'exico}

\author{Sujoy K. Modak}
\email{smodak@ucol.mx}
\affiliation{Facultad de Ciencias - CUICBAS, Universidad de Colima, Colima, C.P. 28045, M\'exico}
\affiliation{Department of Physics, California State University, Fresno, 2345 E San Ramon Ave., M/S MH37, Fresno CA 93740-8031, United States}
\date{October 26, 2021}

\begin{abstract}
    We study some physical aspects of quantum field theory in a two stage universe starting from the inflationary de Sitter and transiting into the radiation dominated stage. We look into the time evolution of the primordial vacuum states, associated with the (i) comoving and (ii) Bunch-Davies modes. We show how the power spectrum for a comoving observer, obtained from the excitation of the aforementioned states defined in the de Sitter stage, changes as the universe transits  into the radiation stage. In addition, we also develop a methodology to transfer the well known result of particle creation in the static de Sitter frame, originating from the aforementioned vacuum states, while the universe makes a transition to the next (radiation dominated) stage.
    {\footnote{\bf{This work is dedicated to the fond memory of Professor Thanu Padmanabhan.}}}
\end{abstract}

\maketitle

\tableofcontents

\section{Introduction}
Quantum Field Theory in Curved Spacetime (QFTCS), often considered to be a low energy limit of any viable theory of ``quantum gravity'', is a set up, where all matter fields are treated quantum mechanically, but the background spacetime has a classical description, such as, using general relativity. Naturally, studies of quantum fields in the context of cosmology would require such a description. The relevant spacetime representing the expanding universe can be expressed using many coordinates, but the natural one, that satisfies our observation of the homogeneous and isotropic universe at large scales, is the comoving frame. This standard cosmological model is based on the FLRW (Friedmann-Lemaitre-Robertson-Walker) metric, which,  following the Standard Model of cosmology, has passed some distinct stages of expansions in which the rates of expansion were very different. The four principal stages in which the universe has evolved so far are believed to be: (i) the early inflationary de-Sitter stage, (ii) the radiation dominated stage, followed by (iii) a matter dominated, and finally (iv) a late dark-energy dominated de-Sitter stage (current stage). Our study in this article is focused to understand some aspects of quantum field theory in the two first stages of the early universe (i.e., the de-Sitter stage and the radiation stage). We want to study how does the primordial vacuum states, defined at the beginning of early de-Sitter stage, can possibly be affected by the evolution of the universe as well as by the transition of the same. In addition, we shall also study the related physical aspects of quantum field theory that could be understood by an interplay between the static and comoving system of coordinates, defined in the first two stages.

In the literature there exists a wide range of articles which are excellent starting points for studying QFTCS \cite{c14}-\cite{cBates}. One important information from these studies were particle creation due to gravitational field which not only makes a spacetime ``curved'' but also disturbs vacuum states of various matter fields. Foundational works on this  and related aspects were laid half a century ago and has a sizable literature  (for an incomplete list, see \cite{c23} -\cite{cHawking}). The most well studied cosmological case, among all four expansion stages, is certainly the de Sitter stage \cite{c4}-\cite{cKim} due its importance for inflation, generation of seeds of structure formation etc.  

In  \cite{c13} some intricate conceptual issues were clarified which were left untouched by earlier studies. It was shown how two different choices vacuum states, given by the Bunch Davies \cite{cBD} and comoving states, in the de Sitter stage, would provide different results for particle creation; added to this was a discussion on how static observers in the de Sitter perceive the Bunch-Davies and comoving vacuums. Physical understanding of the evolution of the power spectrum would depend on the choices of vacuum states which for the Bunch-Davies and comoving vacuums have subtle differences.  In this article, our principal aim is to make a platform to extend the above results beyond the de-Sitter stage. While one would like to extend it for all of the three following epochs, it is technically a daunting task, hence we would consider only the next expansion stage which is dominated by radiation. There are not as many existing works (\cite{c40, Aoki, ijmpd} are some examples of such a study) that are dedicated to study the targeted physical aspects related with the transition of universe from one stage to another, and therefore, extending  the above-mentioned results to the radiation stage is a progressive step in this direction.  

The organization of this article is as follows: In section \ref{QFindSsection}, considering a massless scalar field, we review the calculation of the power spectrum associated with various primordial vacuum states. Then, in section \ref{QFindStoRDsection} we show how the power spectrum, originated from the Bunch-Davies and comoving vacuum states change as the universe transits into the radiation stage. In the next section \ref{TdS}, we review another previous result of particle creation in the static de Sitter frame, both considering the BD and comoving vacuum states. This enables us to look for a methodology for extending the above result in the radiation stage. In order to do so, in section \ref{dsinrad} we extend the static de Sitter coordinates into the radiation stage, i.e., expression of the radiation dominated universe in static de Sitter coordinates. Section \ref{psrd} provides the result of the power spectrum, as perceived by a static de Sitter observer in the radiation stage, due to the excitation of the primordial vacuum states in the earlier epoch (de Sitter). Finally we give a summary of results and conclusions in section \ref{con}.
\vspace{3cm}


\section{Quantum field theory in the de Sitter universe}\label{QFindSsection}

\subsection{Quantum fields in the de Sitter universe}

Consider a free massless field $\Phi(\textbf{x},t)$ in the FRW background. The field equation  is given by
\begin{equation}
    \ddot{\Phi}+3\left(\frac{\dot{a}}{a}\right)\dot{\Phi}-\frac{1}{a^2}\Delta\Phi=0, \label{fieldequation}
\end{equation}
where the dots denote derivatives with respect to the comoving time $t$. As usual, the field can be expanded in terms of a complete set of orthonormal functions $f_\textbf{k}$
\begin{equation}
    \Phi(\textbf{x},t)=\int\frac{d^3k}{(2\pi)^3}\{\hat{a}_\textbf{k}f_\textbf{k}(\textbf{x},t)+\hat{a}_\textbf{k}^\dagger f_\textbf{k}^*(\textbf{x},t)\},\label{fieldexpansiondS}
\end{equation}
with $f_\textbf{k}=e^{i\textbf{k}\cdot\textbf{x}}\psi_k(t)$ (this expression of the field modes is allowed due to spatial homogeneity) and $k=|\textbf{k}|$. Then, from the field equation (\ref{fieldequation}) one finds that $\psi_k(t)$ satisfies the equation
\begin{equation}
    \ddot{\psi}_k+3\left(\frac{\dot{a}}{a}\right)\dot{\psi}_k+\frac{k^2}{a^2}\psi_k=0. \label{psiequation}
\end{equation}
The above equation in the de Sitter metric becomes
\begin{equation}
    \ddot{\psi}_k+3{\cal H}\dot{\psi}_k+\exp(-2{\cal H}t)k^2\psi_k=0.
\end{equation}
and its solution can be written as
\begin{equation}
    \psi_k(t)={\cal A}_ks_k(t)+{\cal B}_ks_k^*(t) \label{generalfieldmodesdS4}
\end{equation}
with
\begin{equation}
    s_k(t)=\frac{1}{\sqrt{2k}}\exp\left[-\frac{ik}{\cal H}(1-e^{-{\cal H}t})\right]\left(\frac{i{\cal H}}{k}+e^{-{\cal H}t}\right).
\end{equation}
Sometimes, it is also useful to treat the scale factor $a$ as the time variable. For such a choice, the functions $s_k(a)$ become
\begin{equation}
    s_k(a)=\frac{1}{\sqrt{2k}}\exp\left[-\frac{ik}{\cal H}\left(1-\frac1a\right)\right]\left(\frac{i{\cal H}}{k}+\frac1a\right).
\end{equation}

\subsection{Choice of vacuum}


As is well known, in a time dependent background, definition of vacuum state is non-trivial. To elaborate, the lowest energy state of the quantum field field, which is interpreted as the  vacuum state, is not time transnational invariant. Indeed, one might invoke the concept of a instantaneous vacuum states defined at every point of time for such a  background. Often, vacuum states are chosen from some physical considerations. In quantum cosmology, there are two most popular choices that has been surfacing over the years. In the following we shall define these two vacuum states and discuss physics related with each of them.

\subsubsection{Bunch-Davies vacuum state}

In the literature \cite{c13} it has been shown that one can choose the mode functions to define a vacuum state in the asymptotic past with respect to the conformal time \footnote{which, in our framework, is related to the comoving time $t$ by $\eta\equiv(1-e^{-{\cal H}t})/{\cal H}$, where the integration constant is chosen in such a way that in the limit ${\cal H}\rightarrow0$, $\eta\rightarrow t$.} where they don't feel the curvature of the spacetime. That is, by requiring that the mode functions behave as the positive frequency modes (identical to those for the Minkowski spacetime) in the limit $\eta\rightarrow-\infty$ (or $k|\eta|\gg1$). This is achieved by imposing this limit on the auxiliary field $\chi\equiv a\Phi$ instead of the physical field $\Phi$. With this choice one must have
\begin{equation}
    {\cal A}_k=1,\,\,\,\,\,\,\,\,{\cal B}_k=0, \label{ABforBD}
\end{equation}
which describes the conventional Bunch-Davies (BD) vacuum. Therefore, the  BD mode functions are then
\begin{equation}
    \psi_k^{(BD)}(t)=s_k(t)=\frac{1}{\sqrt{2k}}\exp\left[-\frac{ik}{\cal H}(1-e^{-{\cal H}t})\right]\left(\frac{i{\cal H}}{k}+e^{-{\cal H}t}\right)
\end{equation}
in the de Sitter stage. The above expression is written in terms of the comoving time but we can express it in terms of the conformal time such that the asymptotic behavior mentioned above is manifest.

\subsubsection{Comoving vacuum state}

Another choice of a vacuum discussed in \cite{c13} is defined by demanding that at an initial time $t=t_0$ the field modes behave like positive frequency modes, with respect to the comoving time $t$. Because of time translation invariance of the metric, one can take $t_0=0$, without lost of generality. Then, it was found that the choice
\begin{equation}
    {\cal A}_k=\frac{\cH+2ik}{2ik},\,\,\,\,\,\,\,\,{\cal B}_k=\frac{\cH}{2ik} \label{ABforCM}
\end{equation}
for the equation (\ref{generalfieldmodesdS4}), provides a definition for the so called ``comoving''  vacuum state for the field modes in the de Sitter phase.

\subsection{Mixing coefficients and power spectrum}

As the comoving time goes by, the expansion of the universe will evolve the mode functions to a mixture of positive and negative frequency modes (simply because the initial vacuum state will no longer remain a vacuum anymore). The procedure to analyze this mixture in \cite{c13} starts by expanding $\psi_k(t)$ in terms of the mixing coefficients $\alpha_\nu$ and $\beta_\nu$ as
\begin{equation}
    \psi_k(t)=\int_0^\infty\frac{d\nu}{2\pi}(\alpha_{\nu k} e^{-i\nu t}+\beta_{\nu k} e^{i\nu t})
\end{equation}
or
\begin{equation}
    \psi_k(t)=\int_{-\infty}^\infty\frac{d\nu}{2\pi}f_k(\nu)e^{-i\nu t}
\end{equation}
so that
\begin{equation}
    \alpha_{\nu k}=f_k(\nu),\,\,\,\,\,\,\,\,\,\beta_{\nu k}=f_k(-\nu);\,\,\,\,\,\,\,\,\,\nu>0.
\end{equation}
In order to determine the mixing coefficients it is needed to calculate the Fourier transform of $\psi_k(t)$
\begin{equation}
    f_k(\nu)=\int_{-\infty}^\infty dte^{i\nu t}\psi_k(t).
\end{equation}

To calculate the last expression we refer to the solution of the time dependent part of the field modes in the de Sitter spacetime (see Eq. (\ref{generalfieldmodesdS4})),
\begin{equation}
    \psi_k(t)={\cal A}_ks_k(t)+{\cal B}_ks_k^*(t)=\psi_{k,1}(t)+\psi_{k,2}(t).
\end{equation}
Following the procedure adopted in \cite{c13} we can define $f_1(\nu)$ and $f_2(\nu)$ as
\begin{equation}
    f_1(\nu)=\int_{-\infty}^\infty dte^{i\nu t}\psi_{k,1}(t),\,\,\,\,\,\,\,f_2(\nu)=\int_{-\infty}^\infty dte^{i\nu t}\psi_{k,2}(t)
\end{equation}
After integration, these functions are found to be

\begin{equation}
f_1(\nu)={\cal A}_k\frac{2e^{-ik/\cH}e^{\pi\nu/2\cH}}{(2k)^{3/2}}(k/\cH)^{i\nu/\cH}\Gamma({-i\nu}{\cH})(i+\nu/\cH) \label{f1}
\end{equation}
and
\begin{equation}
    f_2(\nu)={\cal B}_k\frac{2e^{ik/\cH}e^{-\pi\nu/2\cH}}{(2k)^{3/2}}(k/\cH)^{i\nu/\cH}\Gamma(-i\nu/\cH)(-i-\nu/\cH), \label{f2}
\end{equation}

and $f_k(\nu)=f_1(\nu)+f_2(\nu)$.

\subsubsection{Power spectrum with Bunch-Davies vacuum}

The power spectrum can be calculated by replacing the corresponding values of ${\cal A}_k$ and ${\cal B}_k$ in Eq. (\ref{ABforBD}) and computing the modulus square of $f(\pm \nu)$. This gives
\begin{equation}
    \nu|\alpha_{\nu k}|^2=\nu|f_k(\nu)|^2=\frac{{\cal H}^2}{2k^3}\frac{\beta e^{\beta\nu}}{e^{\beta\nu}-1}\left(1+\frac{\nu^2}{{\cal H}^2}\right),
\end{equation}
and
\begin{equation}
    \nu|\beta_{\nu k}|^2=\nu|f_k(-\nu)|^2=\frac{{\cal H}^2}{2k^3}\frac{\beta}{e^{\beta\nu}-1}\left(1+\frac{\nu^2}{{\cal H}^2}\right);\,\,\,\,\,\,\,\,\,\beta=\frac{2\pi}{{\cal H}}.
\end{equation}
Notice that these spectra are not strictly thermal, because there appears factor $(1+\nu^2/{\cal H}^2)$ which shows a slight departure from thermality. This behaviour has close similarity with thermal spectrum but not exactly so. 


\subsubsection{Power spectrum with comoving vacuum}

For the field modes starting from the comoving vacuum with ${\cal A}_k$ and ${\cal B}_k$ given in \eqref{ABforCM}, the power spectrum for negative frequencies $\nu|f_k(-\nu)|^2$ can be found to be
\ber
    \nu|f_k(-\nu)|^2&=&\nu|f_1(-\nu)|^2+\nu|f_2(-\nu)|^2+2\nu|f_1(-\nu)||f_2(-\nu)|\cos{\theta}\nonumber\\
    &=&\frac{{\cal H}^2\beta}{2k^3}\left(1+\frac{\nu^2}{{\cal H}^2}\right)\left[\left(1+\frac{{\cal H}^2}{4k^2}+\frac{{\cal H}^2}{4k^2}e^{\beta\nu}\right)N\right.\nonumber\\
    &&\left.+\frac{\cal H}{k}\sqrt{1+\frac{{\cal H}^2}{4k^2}}\sqrt{N(N+1)}\cos{\theta}\right],\label{PSCM}
\eer
where $\theta=\arg(f_1,f_2)$ and $N$ is the Planckian factor
\begin{equation}
    N=\frac{1}{e^{\beta\nu}-1}.
\end{equation}

\section{Quantum fields in transit from the de Sitter to the radiation stage}\label{QFindStoRDsection}

\subsection{The de Sitter to radiation transition}

As it was discussed in the previous sections, the FRW metric in comoving frame (for the $k=0$ model) is written as
\begin{equation}
ds^2 = dt^2 - a^2(dr^2 + r^2 d\Omega^2) \label{com}
\end{equation}
We are considering the phase where the early universe starts as the inflationary stage and transits into the radiation stage. The scale factors for these two stages and their derivatives can be made equal at the transition point, giving
\begin{eqnarray}
a(t) &=& e^{{\cal H} t}   \hspace{13pt} (t \le t_r) \label{aoft1}\\
        &&(2{\cal H} e)^{1/2}\, t^{1/2}  \hspace{12pt} (t\geq t_r )
\label{aoft}
\end{eqnarray}
where $t_r = 1/2{\cal H}$ is the transition time.  It is important to mention that the Hubble parameter $H(t)=\dot{a}/a$ is just a constant in the de Sitter stage, given by ${\cal H}$, but $H(t)=\dot{a}/a$ becomes time dependent in the radiation stage, given by $H=1/2t$.

\subsection{Conformal diagram of a universe with the  de Sitter and radiation stages}
\begin{figure}
    \centering
    \includegraphics[trim={12.4cm 3cm 30cm 8cm},clip,scale=0.5]{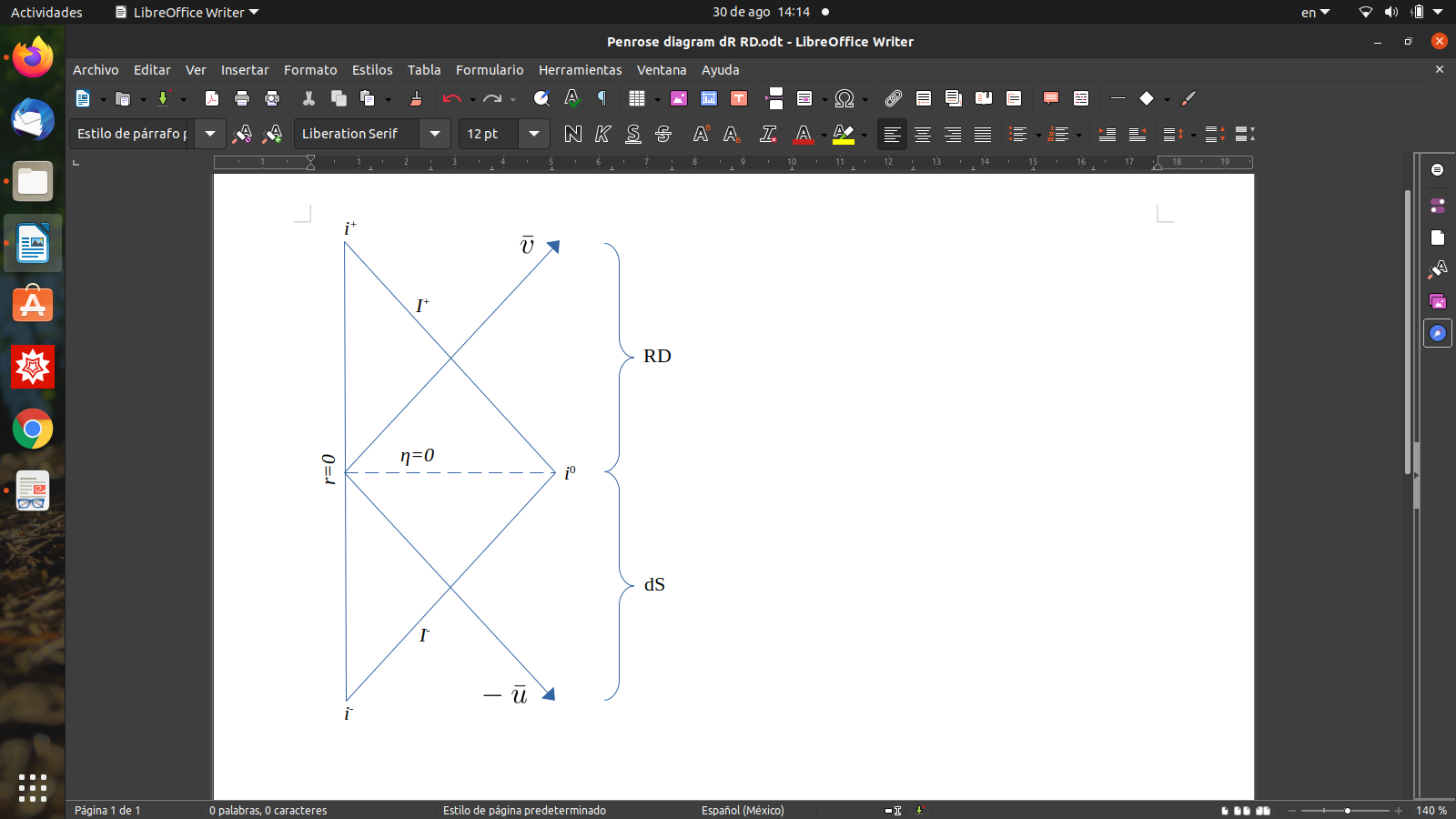}
    \caption{Penrose diagram of a de Sitter universe transiting to the radiation stage.}
    \label{penrosedSRD}
\end{figure}

One way to construct the Penrose diagram of the de Sitter transiting into the radiation dominated spacetime is by considering the conformally flat form of the de Sitter metric
\be
ds^2=a(\eta)^2(d\eta^2-dr^2-r^2d\Omega^2)\label{99}
\ee
where $\eta$ is the conformal time, defined as
\be
\eta=\int_t^{1/2\cH}\f{dt}{e^{\cH t}}=\f{1}{\cH}\left(e^{-\cH t}-e^{-1/2}\right)
\ee
for the de Sitter stage and
\be
\eta=\int_{1/2\cH}^t\f{dt}{(2\cH et)^{1/2}}=\sqrt{\f{2}{\cH e}}\left(t^{1/2}-\f{1}{\sqrt{2\cH}}\right)
\ee
for the radiation stage.
In this conformally flat representation, one can follow the standard procedure to construct the Penrose diagram of the Minkowski spacetime. Consider the null coordinates
\be
u=\eta-r,\,\,\,\,\,\,\,\,v=\eta+r.
\ee
The metric becomes
\be
ds^2=a(u,v)^2\left[dudv-\left(\f{v-u}{2}\right)^2d\Omega^2\right].
\ee
Now, let's consider the following transformation
\be{}
\bar{u}\tan^{-1}{u},\,\,\,\,\,\,\,\,\bar{v}=\tan^{-1}{v}.
\ee
The metric becomes
\be
ds^2=\f14 a(\bar{u},\bar{v})^2\sec^2{\bar{u}}\sec^2{\bar{v}}[4d\bar{u}d\bar{v}-\sin^2(\bar{v}-\bar{u})d\Omega^2].
\ee
In figure \ref{penrosedSRD} we can see the penrose diagram of the whole spacetime constructed from the conformally flat metric \eqref{99}.
Notice that the transition phase at $t=t_r$ correspnds to the $\eta=0$ line (dashed line).

\subsection{Field modes in the radiation dominated universe}

As we previously discussed, the field equation for a massless scalar field in the FRW background reads
\begin{equation}
    \ddot{\Phi}+3\left(\frac{\dot{a}}{a}\right)\dot{\Phi}-\frac{1}{a^2}\Delta\Phi=0 
\end{equation}
Consider again the field expansion
\begin{equation}
    \Phi(\textbf{x},t)=\int\frac{d^3k}{(2\pi)^3}\{\hat{a}_\textbf{k}f^R_\textbf{k}(\textbf{x},t)+\hat{a}_\textbf{k}^\dagger f^{R*}_\textbf{k}(\textbf{x},t)\},\label{fieldexpansionRD}
\end{equation}
where $\hat{a}_{\textbf{k}}$ and $\hat{a}_{\textbf{k}}^\dagger$ have been taken to be the same creation and annihilation operators as in de Sitter, and $f^\text{R}_\textbf{k}=e^{i\textbf{k}\cdot\textbf{x}}\psi_k^\text{R}(t)$, so that $\psi_k^\text{R}(t)$ satisfies the equation
\begin{equation}
    \ddot{\psi}_k^\text{R}+3\left(\frac{\dot{a}}{a}\right)\dot{\psi}_k^\text{R}+\frac{k^2}{a^2}\psi_k^\text{R}=0, \label{psiRequation}
\end{equation}
which, for $a=(2{\cal H}e)^{1/2}t^{1/2}$, becomes
\begin{equation}
    \ddot{\psi}_k^\text{R}+\frac{3}{2t}\dot{\psi}_k^\text{R}+\frac{k^2}{2{\cal H}et}\psi_k^\text{R}=0.
\end{equation}

The solution of the last equation can be written as
\begin{equation}
    \psi_k^\text{R}(t)=\frac{1}{2\sqrt{k{\cal H}et}}\left[C_k\exp\left(ik\sqrt{\frac{2t}{{\cal H}e}}\right)+D_k\exp\left(-ik\sqrt{\frac{2t}{{\cal H}e}}\right)\right] \label{psiR(t)}
\end{equation}
for $t>0$ or, using $a$ instead of $t$ as the time variable,
\begin{equation}
    \psi_k^\text{R}(a)=\frac{1}{\sqrt{2k}a}\left(C_ke^{ika/e{\cal H}}+D_ke^{-ika/e{\cal H}}\right). \label{psiR(a)}
\end{equation}

At the transition point (at $t=t_r=1/2{\cal H}$) the field and its comoving time-derivative must transit smoothly. Then, for a given $\textbf{k}$, the functions $\psi_k$ and $\psi_k^\text{R}$ and their derivatives must satisfy the conditions
\begin{equation}
    \psi_k(a_r)=\psi_k^\text{R}(a_r),\,\,\,\,\,\,\,\,\,\,\psi_k(a_r)'=\psi_r^\text{R}(a_r)' \label{matchcond}
\end{equation}
where the prime denotes derivative with respect to $a$ and $a_r=a(t_r)=e^{1/2}$.

\subsection{Choice of vacuum}

According to the choice of vacuum set in the de Sitter stage one can determine coefficients $C_k$ and $D_k$ in expression (\ref{psiR(a)}) from the matching conditions (\ref{matchcond}).
If one works with the Bunch-Davies vacuum choice,  these coefficients  are found to be
\begin{equation}
    C_k^{\text{(BD)}}=\frac{{\cal H}^2}{2k^2}e^{1-ik/{\cal H}} \label{CkBD}
\end{equation}
and
\begin{equation}
    D_k^{\text{(BD)}}=\frac12e^{-ik(1-2e^{-1/2})/{\cal H}}(-e{\cal H}^2/k^2+2ie^{1/2}{\cal H}/k+2). \label{DkBD}
\end{equation}
We can also compute these coefficients for the field modes in the radiation stage starting from a comoving vacuum (in the de Sitter stage). A straight-forward calculation yields
\ber
    C_k^{\text{(CM)}}&=&\frac{e^{ik(1-2e^{-1/2})/{\cal H}}{\cal H}^3}{4k^3}\left[ie-2e^{1/2}k/{\cal H}\right.\nonumber\\
    &&\left.-2ik^2/{\cal H}^2+e^{1-2ik(1-e^{-1/2})/{\cal H}}(-i+2k/{\cal H})\right] \label{CkCM}
\eer
and
\ber
    D_k^{\text{(CM)}}&=&\frac{e^{ik(1+2e^{-1/2})/{\cH}}\cH^3}{4k^3}\left[-ie^{1-2ike^{-1/2}/\cH}\right.\nonumber\\
    &&\left.+2e^{1/2-2ik/\cH}(1+2ik/\cH)k/\cH+e^{-2ik/\cH}(2k^2/\cH^2-e)(-i+2k/\cH)\right]. \label{DkCM}
\eer
With these coefficients we now have a complete information of the field modes which started from the de Sitter with different initial conditions, both are physical in their own sense, and then evolved into the radiation stage smoothly passing the transition point.

\subsection{Mixing coefficients and power spectrum}

We can extend the previous results for the power spectrum, due to the comoving and Bunch-Davies choice of vacuum states, into the radiation dominated stage. This calculation is new.    

In the radiation stage, by following the same procedure as for the de Sitter case, we can expand the field modes, $\psi_k^\text{R}(t)$, as
\begin{equation}
    \psi_k^\text{R}(t)=\int_0^\infty\frac{d\nu}{2\pi}(\alpha_\nu^R e^{-i\nu t}+\beta_\nu^R e^{i\nu t})
\end{equation}
or
\begin{equation}
    \psi_k^\text{R}(t)=\int_{-\infty}^\infty\frac{d\nu}{2\pi}g(\nu)e^{-i\nu t} \label{expansionofpsiR}
\end{equation}
so that $\alpha_\nu^\text{R}=g(\nu)$ and $\beta_\nu^\text{R}=g(-\nu)$ with $\nu>0$. Thus, the inverse Fourier transform of (\ref{expansionofpsiR}) is
\begin{equation}
    g(\nu)=\int_{-\infty}^\infty dte^{i\nu t}\psi_k^\text{R}(t). \label{g}
\end{equation}
Since the scale factor of the radiation dominated universe is only well defined for $t>0$, the field modes are also well defined only for $t>0$ (or $a>0$) (therefore the $t<0$ region has a vanishing measure). Now, let's define $g_1(\nu)$ and $g_2(\nu)$ as the Fourier transforms of the first and second terms of $\psi_k^\text{R}(t)$ in eq. (\ref{psiR(t)}) respectively as
\begin{equation}
    g_1(\nu)=\int_0^\infty dte^{i\nu t}\frac{1}{2\sqrt{k{\cal H}et}}C_ke^{ik\sqrt{2t/{\cal H}e}}, \label{g1}
\end{equation}
and
\begin{equation}
    g_2(\nu)=\int_0^\infty dte^{i\nu t}\frac{1}{2\sqrt{k{\cal H}et}}D_ke^{-ik\sqrt{2t/{\cal H}e}}. \label{g2}
\end{equation}

We can use $a$ instead of $t$ as the time variable and take the above $t$-integrals to $a$-integrals with the changes $a=(2eH)^{1/2}t^{1/2}$ and $dt=(a/eH)da$. Then these integrals could be solved for, giving
\begin{eqnarray}
    g_1(\nu)&=&\int_0^\infty da\frac{1}{eH}e^{i\nu a^2/2eH}\frac{1}{\sqrt{2k}}C_ke^{ika/eH}\\
    &=&\f{C_ke^{-1/2-ik^2/2e\cH\nu}\sqrt{\pi}}{2\sqrt{-ik\cH\nu}}\left[1-\erf\left(k\sqrt{\f{-i}{2e\cH\nu}}\right)\right]
\end{eqnarray}
and
\begin{eqnarray}
    g_2(\nu)&=&\int_0^\infty da\frac{1}{eH}e^{i\nu a^2/2eH}\frac{1}{\sqrt{2k}}D_ke^{-ika/eH}\\
    &=&\f{D_ke^{-1/2-ik^2/2e\cH\nu}\sqrt{\pi}}{2\sqrt{-ik\cH\nu}}\left[1+\erf\left(k\sqrt{\f{-i}{2e\cH\nu}}\right)\right].
\end{eqnarray}

Then
\begin{equation}
    |g_1(\nu)|^2=\f{|C_k|^2\pi}{4ke\cH\nu}\left[1-\erf\left(k\sqrt{\f{i}{2e\cH\nu}}\right)\right]\left[1-\erf\left(k\sqrt{\f{-i}{2e\cH\nu}}\right)\right],
\end{equation}
and
\begin{equation}
    |g_2(\nu)|^2=\f{|D_k|^2\pi}{4ke\cH\nu}\left[1+\erf\left(k\sqrt{\f{i}{2e\cH\nu}}\right)\right]\left[1+\erf\left(k\sqrt{\f{-i}{2e\cH\nu}}\right)\right].
\end{equation}

\begin{figure}
    \centering
    \includegraphics[scale=0.7]{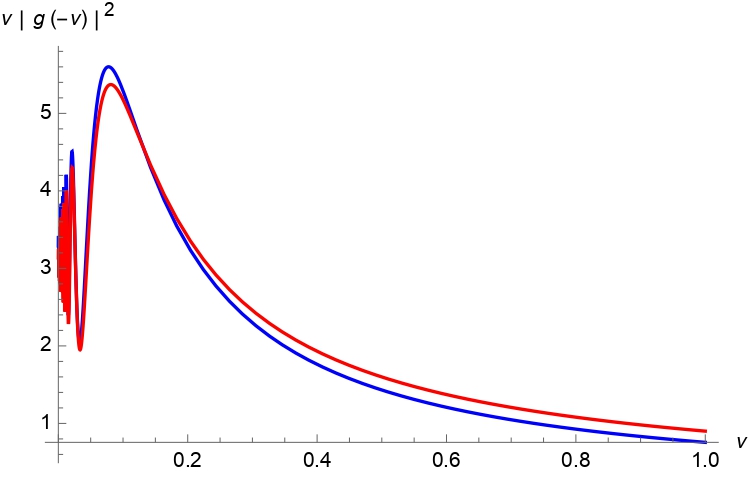}
    \caption{Power spectrum of the field evolved from Bunch-Davies vacuum (blue graph) and from comoving vacuum (red graph) in the radiation stage with $k=\cH=1$.}
    \label{bothpsRD}
\end{figure}

\begin{figure}[h]
    \centering
    \includegraphics[scale=0.7]{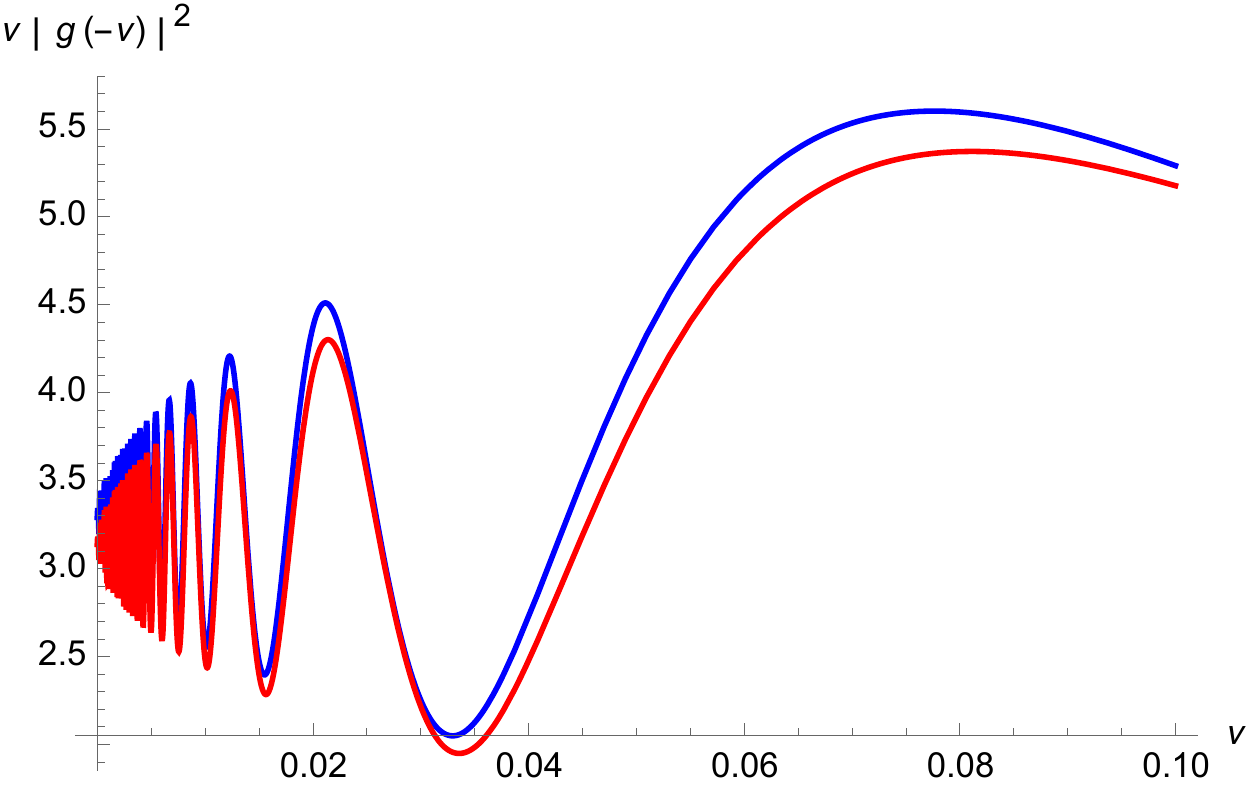}
    \caption{Zoom in of the oscillating part of Figure 2. These oscillations are nonexistent in the de Sitter stage which has a thermal or near thermal power-spectrum.}
    \label{bothpsRDzoom}
\end{figure}

For negative frequencies, we take $\nu$ to $-\nu$ in the expressions for $g_1(\nu)$ and $g_2(\nu)$; this gives
\begin{equation}
    |g_1(-\nu)|^2=|g_1(\nu)|^2,\,\,\,\,\,\,\,\,\,\,|g_2(-\nu)|^2=|g_2(\nu)|^2.
\end{equation}
Therefore the complete power spectrum is given by
\begin{equation}
    \nu|\beta_\nu^\text{R}|^2=\nu|g(-\nu)|^2=\nu|g_1(-\nu)|^2+\nu|g_2(-\nu)|^2+2\nu|g_1(-\nu)||g_2(-\nu)|\cos{\theta^\text{R}}
\end{equation}
where $\theta^\text{R}=\arg(g_1,g_2)$.

Finally, combining all the above expressions of modulus square we can determine
\ber
    \nu|g(-\nu)|^2&=&\f{|C_k|^2\pi}{4ke\cH}\left[1-\erf\left(k\sqrt{\f{i}{2e\cH\nu}}\right)\right]\left[1-\erf\left(k\sqrt{\f{-i}{2e\cH\nu}}\right)\right]\nonumber\\
    &&+\f{|D_k|^2\pi}{4ke\cH}\left[1+\erf\left(k\sqrt{\f{i}{2e\cH\nu}}\right)\right]\left[1+\erf\left(k\sqrt{\f{-i}{2e\cH\nu}}\right)\right]\nonumber\\
    &&+\f{|C_k||D_k|\pi}{2ke\cH}\sqrt{\left[1-\erf^2\left(k\sqrt{\f{i}{2e\cH\nu}}\right)\right]\left[1-\erf^2\left(k\sqrt{\f{-i}{2e\cH\nu}}\right)\right]}\cos{\theta^R}.
    \label{psrd00}
\eer
We see now the problem is solved: we just need to consider which vacuum state are we interested in and chose the pair of coefficients ($C_k$ and $D_k$) from the corresponding expressions \eqref{CkBD} and \eqref{DkBD} or \eqref{CkCM} and \eqref{DkCM}.
This will provide us desired result of how the power spectrum looks like when the primordial vacuum states (i.e., BD and comoving vacuums defined in the de Sitter stage) are excited due to the expansion of the universe. 

In Figs. \ref{bothpsRD} and \ref{bothpsRDzoom} we show the behavior of the power spectrum for the field modes evolved from the Bunch-Davies and comoving vacuum states. The difference in the peak and period of each  oscillations are detectable for these cases. Such an oscillatory behavior for low and intermediate frequencies is a direct signature of the transitions between the expansion stages, which eventually dies out for large frequencies.



\section{Power spectrum in the de Sitter coordinates}\label{TdS}

One important aspect that was discussed in \cite{c13} was the calculation of the power spectrum (particle creation) for static de Sitter observer looking into the comoving and Buch-Davies vacuum states. Here we give a brief review odf the analysis.

In comoving coordinates the de Sitter metric reads
\begin{equation}
    ds^2=dt^2-e^{2{\cal H}t}(dr^2+r^2d\Omega^2).
\end{equation}
Using the transformation
\begin{equation}
    R=e^{{\cal H}t}r;\,\,\,\,\,\,\,\,T=t-\frac{1}{2\cal H}\ln(1-{\cal H}^2R^2)
\end{equation}
and defining the tortoise coordinate
\begin{equation}
    R_*=\int\frac{dR}{1-{\cal H}^2R^2}
\end{equation}
the de Sitter metric takes the static form 
\ber
ds^2 &=& \left(1-\cH^2R^2\right)dT^2-\frac{dR^2}{\left(1-\cH^2R^2\right)}-R^2d\Omega^2\\
&=& \left(1-{\cH}^2R^2(R_*)\right)\left(dT^2-dR_*^2\right)-R^2(R_*)d\Omega^2.
\eer
The field equation in static coordinates becomes
\begin{equation}
    \left[\frac{\partial^2}{\partial T^2}-\frac{f(R)}{R^2}\frac{\partial}{\partial R}\left(R^2f(R)\frac{\partial}{\partial R}\right)-\frac{f(R)\hat{L}^2}{R}\right]\Phi(T,R,\Omega)=0
\end{equation}
where $f(R)\equiv1-{\cal H}^2R^2$ and $\hat{L}$ is the angular Laplacian operator. Consider $\Phi=\phi_l(R)Y_{lm}(\Omega)e^{-i\omega T}/R$. Then, one finds
\begin{equation}
    -\omega^2\phi_l-\frac{f}{R}\frac{d}{dR}\left(R^2f\frac{d}{dR}\left(\frac{\phi_l}{R}\right)\right)-\frac{l(l+1)f}{R^2}\phi_l=0.
\end{equation}
Notice that $f(R)\rightarrow0$ near the horizon $R\rightarrow1/{\cal H}$. Thus, the dominant contribution comes from the $s$-mode near the horizon. Then, focusing on $l=0$ mode, $\phi$ satisfies the equation
\begin{equation}
    \frac{d^2\phi}{dR_*^2}+\left(\omega^2-\frac{ff'}{R}\right)\phi=0,
\end{equation}
where $f'=df/dR$. It is easy to see that in the near the horizon ($f\rightarrow0$) the solutions $\phi$ behave as $\exp(\pm i\omega R_*)$ and therefore, in the past horizon limit ($R\rightarrow1/{\cal H},T\rightarrow-\infty$, the modes in the static system behave as
\be
\phi_\text{st}=\frac{e^{\pm i\omega V}}{\sqrt{2\omega}}\label{phist}
\ee
where $V=T+R_*$.

On the other hand, the Bunch-Davies vacuum modes in spherical coordinates can be expressed as
\begin{equation}
    \Phi_k^\text{BD}=\frac{e^{-ik/{\cal H}}}{\sqrt{2k}}\sum_{l=0}^\infty i^l(2l+1)j_l(kr)P_l(\cos{\theta})e^{ik/{\cal H}e^{-{\cal H}t}}\left(\frac{i{\cal H}}{k}+e^{-{\cal H}t}\right).\label{BDspherical}
\end{equation}
Considering only the $s-$wave contribution and expressing the above expression in ($T,R_*$) coordinates, we have
\begin{equation}
    \Phi_k^\text{BD}=\frac{e^{-ik/{\cal H}}}{\sqrt{2k}}\left(e^{ik/{\cal H}e^{-{\cal H}U}}-e^{ik/{\cal H}e^{-{\cal H}V}}\right)\left(\frac{i{\cal H}}{k}e^{{\cal H}T}(1-{\cal H}^2R^2)+1\right)
\end{equation}
which, in the past horizon limit ($R\rightarrow1/{\cal H},T\rightarrow-\infty$, gives
\begin{equation}
    \Phi_k^\text{BD}\rightarrow\frac{1}{\sqrt{2k}}e^{i(k/{\cal H})e^{-{\cal H}V}}\label{phist2}.
\end{equation}

The Bogoliubov coefficients that relate the modes \eqref{phist} and \eqref{phist2} are
\begin{equation}
    \beta_{\omega k}=-i\int_T dR_*(\phi_\text{st}\partial_T\Phi_k^\text{BD}-\Phi_k^\text{BD}\partial_T\phi_\text{st}).\label{KGinnerproduct}
\end{equation}
Since the Klein-Gordon inner product is independent of the surface over which the integral is evaluated, it is convenient to evaluate it on a space-like surface near to the past horizon. The above integral can be written as a $V$-integral; since $R_*=V-T$ then $dR_*=dV$ for fixed $T$ and $\partial_T=\partial_V$:
\begin{equation}
    \beta_{\omega k}=\frac{-i}{\sqrt{2\omega}}\int_{-\infty}^\infty dV(e^{-i\omega V}\partial_V\Phi_k^{BD}(V)-\Phi_k^{BD}(V)\partial_Ve^{-i\omega V}).
\end{equation}
Integrating by parts the first term one gets
\begin{eqnarray}
\beta_{\omega k} &=& \frac{-i}{\sqrt{2\omega}}\left[e^{-i\omega V}\Phi_k^{BD}(V)|_{-\infty}^\infty+2i\omega\int_{-\infty}^\infty dVe^{-i\omega V}\Phi_k^{BD}(V)\right] \\
        &=& \sqrt{2\omega}\int_{-\infty}^\infty dVe^{-i\omega V}\Phi_k^{BD}(V).\label{betaomegak}
\end{eqnarray}
This gives
\begin{eqnarray}
\beta_{\omega k} &=& \sqrt{\frac{\omega}{k}}\int_{-\infty}^\infty dVe^{-i\omega V}e^{i(k/{\cal H})e^{-{\cal H}V}} \\
        &=& \sqrt{\frac{\omega}{k}}\left(\frac{1}{\cal H}\right)\left(\frac{k}{\cal H}\right)^{i\omega/{\cal H}}\Gamma\left(-\frac{i\omega}{\cal H}\right)e^{-\pi\omega/2{\cal H}}.
\end{eqnarray}
The modulus square $|\beta_{\omega k}|^2$ is Planckian at temperature ${\cal H}/2\pi$
\begin{equation}
    |\beta_{\omega k}|^2=\frac{\beta}{k(e^{\beta\omega}-1)};\,\,\,\,\,\,\beta=\frac{2\pi}{\cal H}
\end{equation}
Thus, the Bunch-Davies vacuum has a thermal character when perceived by a static observer. 

The relation to the comoving vacuum can also be obtained by using the fact that
\begin{equation}
    \psi_k^\text{CM}(t)=A_k\psi_k^\text{BD}(t)+B_k\psi_k^{*\text{BD}}(t)
\end{equation}
with ${\cal A}_k$ and ${\cal B}_k$ given in (\ref{ABforCM}). Thus
\begin{equation}
    \Phi_k^\text{CM}\rightarrow\frac{1}{\sqrt{2k}}\left(A_ke^{i(k/{\cal H})e^{-{\cal H}V}}+B_ke^{-i(k/{\cal H})e^{-{\cal H}V}}\right).
\end{equation}
Then the spectrum is
\begin{equation}
    |\beta_{\omega k}|^2=\frac{\beta}{k}\left[(|A_k|^2+|B_k|^2e^{\beta\omega})N+|A_k||B_k|\sqrt{N(N+1)}\cos{\theta}\right]
\end{equation}
where
\begin{equation}
    N=\frac{1}{e^{\beta\omega}-1}
\end{equation}
where again an interference term of the form $\sqrt{N(N+1)}$ has appeared for the comoving vacuum (similar to the power spectrum (\ref{PSCM})) when perceived by static observers.

\section{Transfer of static de Sitter coordinates to the radiation stage}
\label{dsinrad}

Here our goal is to extend the static de Sitter coordinates into the radiation stage. We start from the   static de Sitter spacetime and then make it smoothly transit into the radiation stage where the spacetime is still described using the de Sitter coordinates, which, however, do not remain static anymore. This exercise is necessary if we want to express physics from the viewpoint of the static de Sitter observer in a domain when the universe has already made a transition to the next stage.


Introducing $R = a(t) r$ and expressing the standard FRW metric in $(R,t)$ coordinates, we find
\be
ds^2 = (1- H^2 R^2) dt^2 - dR^2 + 2H R dR dt - R^2 d\Omega^2,
\ee
where $H = (1/a) {da}/{dt}${\footnote{note that $H=constant=\cH$ for the static de Sitter case while it is a time dependent parameter in the radiation stage.}}.
We then introduce a new ``time"
\be
d{\tilde t} = 1/F (dt + b dR) \label{dtildet}
\ee
with $b = \frac{HR}{1-H^2R^2}$ (it ensures that there will not be a $d{\tilde t}dR$ term in the metric) and we need find $F$ 
such that $d{\tilde t}$ is a proper differential satisfying
\be
\partial_{R}(1/F) = \partial_{t}(b/F).
\label{ed}
\ee
This transformation makes \eqref{com} to take the following static form
\be
ds^2 = A F^2 d{\tilde t}^2 - \frac{1}{A} dR^2 - R^2 d\Omega^2
\label{mt1}
\ee
where
\be
A = 1 - H^2R^2.
\ee

It is tricky to find an appropriate expressions for $F$ in the inflationary de Sitter (dS) and radiation dominated (RD) stages in a manner that $F_{dS}$ transits to $F_{RD}$ smoothly at the transition point. If we can ensure this, then  we shall have a universal description of inflationary dS and radiation dominated stages under the umbrella of the static coordinates. We find it by an educated guess and consistency check \eqref{ed} on $F$. The final expressions are the following
\ber
F_{dS} (t,R) = \frac{e^{2\cH t}}{1 - \cH^2 R^2} \label{fds}\\
F_{RD} (t,R) = \frac{e}{1 - H(t)^2 R^2}. \label{frd}
\eer
Note that, in the above expressions we have $\cH$ as a constant (Hubble constant of inflationary universe) and $H(t)$ as the Hubble parameter in radiation stage which depends on the comoving time (i.e., on both $R$ and ${\tilde t}$). As we saw, the transition time from the initial de Sitter to radiation stage can be calculated by equating the scale factor and its derivative at the transition point, which turns out to be $t_{r} = 1/2\cH$. Since at the transition time we  $\cH = H(t_r)$ we have $F_{dS} (t_r = 1/2\cH) = F_{RD} (t_r = 1/2\cH)$.

The integral form of \eqref{dtildet} can now be expressed for the de Sitter and radiation stages just by using  \eqref{fds} and \eqref{frd} and these are given by
\ber
{\tilde t} (t,R) &=& {\tilde t}_{dS} = \frac{1}{2}(\cH R^2 - \frac{1}{\cH}) e^{-2\cH t},\,\,\, t\le 1/2\cH; \label{tildet(t,R)dS}
\\
{\tilde t} (t,R) &=& {\tilde t}_{rd} =\frac{1}{e}(t + \frac{R^2}{4t} - 1/\cH),\,\,\, t\ge 1/2\cH \label{tildet(t,R)RD}
\eer
which in terms of pure comoving coordinates are 
\ber
{\tilde t}_{dS} = \frac{\cH r^2}{2} - \frac{1}{2\cH} e^{-2\cH t}, \hspace{1cm}	t\le 1/2\cH;\\
{\tilde t}_{rd} = \frac{\cH r^2}{2} + \frac{1}{e}(t - 1/\cH), \hspace{1cm} t\ge 1/2\cH.
\eer
It is interesting to check that when expressed in terms of $({\tilde t},R)$ coordinates $F_{dS}$ is independent of $R$
\be
F_{dS} = - \frac{1}{2\cH {\tilde t}}.
\ee
Substituting this in \eqref{mt1} we find the de Sitter spacetime represented by the interval
\be
ds^2 = \frac{1 - \cH^2 R^2}{4\cH^2} \frac{d{\tilde t}^2}{{\tilde t}^2} - \frac{dR^2}{1 - \cH^2 R^2} - R^2 d\Omega^2,
\ee
which can be easily be taken to be the static de Sitter universe
\be
ds^2 = (1 - \cH^2 R^2) {d{T}^2} - \frac{dR^2}{1 - \cH^2 R^2} - R^2 d\Omega^2,
\label{mt2}
\ee
just by a redefinition of time
\be
{\tilde t}=\frac{1}{\cH}e^{2\cH T}. 
\label{red}
\ee
Therefore, we conclude $T,R$ are static de Sitter time and space coordinates.

Now we can also relate the static coordinates with the comoving coordinates,
\ber
T_{\text{dS}} &=& \frac{1}{2\cH}\log\left(\frac{\cH^2 r^2}{2} - \frac{1}{2} e^{-2\cH t} \right) ;\,\,\ R = a_{\text{dS}}(t)r,\,\,\, \text{for}~~ t\le 1/2\cH \label{tr0}\\
T_{\text{RD}} &=& \frac{1}{2\cH}\log\left(\frac{\cH^2 r^2}{2} + \frac{1}{e} (\cH t-1)  \right); \,\,\ R = a_{\text{rd}}(t)r,\,\,\, \text{for}~~t\ge 1/2\cH \label{tr1}.
\eer
Notice that the past horizon is given by the limit $T\rightarrow - \infty$ and $R=1/\cH$.

On the other hand, it is now easy to find the interval, that connects \eqref{mt2}, for the radiation stage. We just need to plug \eqref{frd} in \eqref{mt1} to obtain
\be
ds^2 =  \frac{e^2 d{\tilde t}^2 - dR^2}{1 - H^2 R^2} - R^2 d\Omega^2.
\label{mt3}
\ee
This interval can now be finally expressed using the $T,R$ coordinates by the help \eqref{red}, which yields
\be
ds^2 =  \frac{4 e^2 e^{4\cH T}dT^2 - dR^2}{1 - H^2 R^2} - R^2 d\Omega^2.
\label{mt4}
\ee
Therefore, we conclude that the inflationary static de Sitter spacetime \eqref{mt2} smoothly transits into \eqref{mt4} in the radiation stage and therefore we get a representation of the radiation stage in terms of the static de Sitter coordinates. Notable, static de Sitter coordinates do not remain static in the radiation stage where the metric is dynamical (i.e., function of $T$ as well as $R$). 

\section{Power spectrum  in the extended static de-Sitter frame in the radiation stage}
\label{psrd}

We have already shown, for the de Sitter stage, that the Bunch-Davies vacuum, when viewed from the static de Sitter frame, it is perceived as a thermal state. On the other hand, the comoving vacuum is not really thermal; it includes correlations. In this section we want to see how do these expressions get altered when the universe falls into the radiation stage.

As we have showed in the last section, the static de Sitter coordinates can be smoothly transferred into the radiation dominated spacetime where they are no longer static. They also have a different relationship with the comoving coordinates. These relationships \eqref{tr0} and \eqref{tr1}, in the de Sitter and radiation stages, however, do match at the transition point. 

\subsection{Mode functions in the radiation stage}

First, let us write down mode functions in the comoving coordinates.  The field modes for the radiation stage in the comoving frame were found to be $f^\text{R}_\textbf{k}=e^{i\textbf{k}\cdot\textbf{x}}\psi_k^\text{R}(t)$ with $\psi_k^\text{R}(t)$ given by \eqref{psiR(t)}. We need to use the spherically symmetric decomposition of the plane wave modes to write the mode functions as,
\ber
f^\text{R}_k &=& \sum_{l=0}^\infty i^l(2l+1)j_l(kr)P_l(\cos{\theta})\psi_k^R(t)\nonumber\\
&=& \frac{1}{2\sqrt{k{\cal H}et}}\sum_{l=0}^\infty i^l(2l+1)j_l(kr)P_l(\cos{\theta})\left[C_k\exp\left(ik\sqrt{\frac{2t}{{\cal H}e}}\right)+  \right. \nonumber \\
&& \left. D_k\exp\left(-ik\sqrt{\frac{2t}{{\cal H}e}}\right)\right].
\eer
We want to trace back these modes back to the past horizon for which only important contribution is the s-wave contribution. Considering this limit, we get
\ber
f^\text{R}_k (t, r)&=& \frac{1}{2\sqrt{k{\cal H}et}} \frac{\left(e^{ikr}- e^{-ikr} \right)}{2i kr}\left[C_k\exp\left(ik\sqrt{\frac{2t}{{\cal H}e}}\right)  \right. \nonumber \\
&& \left. +D_k\exp\left(-ik\sqrt{\frac{2t}{{\cal H}e}}\right)\right].\\
&=& \frac{-i}{(2k)^{3/2}}\left[C_k \frac{e^{ikv}}{R} - D_k \frac{e^{-ikv}}{R}\right]\nonumber\\
&&- \frac{-i}{(2k)^{3/2}}\left[C_k \frac{e^{iku}}{R} - D_k \frac{e^{-iku}}{R}\right]
\eer
where $v=\eta+r$ and $u=\eta-r$ with $\eta=\sqrt{2t/\cH e}$.

Now we turn our focus on the radiation stage described by the static de Sitter coordinates but extended forcibly, if we may say so, to the radiation stage and given by \eqref{mt4}. Following the footsteps imprinted by our analysis using comoving coordinates we are now in the need of solving field equation with \eqref{mt4} as the background metric. However,  such an attempt is bound to fail, in terms of obtaining a positive and  negative frequency solutions simply because the metric \eqref{mt4} is non-static. This is why we must device an alternative way which could bypass the problem. Fortunately, there is one such path which we discuss below.

There are two important observations that we can make. First, the metric \eqref{mt4} can be expressed as a conformally static, spherically symmetric form simply a redefinition of time where the new time $T_*$ is related to the static time $T$ in the following way
\begin{equation}
   T_* = \frac{e}{{\cH}}e^{2\cH T}.  
   \label{tstar}
\end{equation}
This new time is almost same as $\tilde{t}$ except one factor. In the $(T_*,R)$ coordinates the metric \eqref{mt4} takes the following form\footnote{This metric was also obtained from somewhat different viewpoint in earlier studies \cite{ijmpd, prd, jhep}},
\begin{equation}
  d{s}^2 = \frac{ dT_*^2 -dR^2}{1-H^2R^2} - R^2 d\Omega^2,   \label{statric}
\end{equation}
which is conformally static. This conformal staticity would allow us to solve the field equation in $(T_*,R)$ frame at least for the $s-$wave approximation. The second observation is that the relation between $T$ and $T_*$, following \eqref{tstar}, is simply a redefinition of time without mixing it with the space coordinates. This relation guarantees that the positive frequency mode with respect to $T_*$ will remain positive frequency with respect to $T$. Therefore, vacuum states will be unique between these two frames and therefore particle creation in $(T_*,R)$ frame will also mean a particle creation in $(T,R)$ frame although frequencies will be altered in two frames (reminiscent of the Doppler effect). We want to exploit these two important features for our discussion. We shall elaborate more once  we carryout required calculations.

Considering the symmetry of the background metric \eqref{statric}, we can decompose the scalar field as
\begin{equation}
    \Phi_\Omega = \sum_{l,m}\frac{\Phi^{lm}_\Omega(T_*,R)}{R}Y_{lm}(\theta, \phi).
\end{equation} 
Then, the Klein-Gordon equation is given by,
\begin{equation}
    \Bigg(\frac{\partial^2\Phi^{lm}_\Omega}{\partial T_*^2}- \frac{\partial^2\Phi^{lm}_\Omega}{\partial R^2}\Bigg) + \frac{l(l+1)}{R^2(1-H^2R^2)}\Phi^{lm}_\Omega = 0,
\end{equation}
where we find an effective potential
\begin{equation}
    V_{l} (R,H) = \frac{l(l+1)}{R^2(1-H^2R^2)}.
    \label{pot}
\end{equation}
For any $l\neq 0$ modes, this potential has an interesting behavior -- it diverges at $R=0$ and $R=1/H$, i.e., at the origin of the coordinate system and also at the Hubble radius. While the divergence at the origin is a characteristic of the spherical waves because of the choice of coordinate system, the other divergence at $R=1/H$ makes our equation nontrivial -- it means that the mode functions with $l\ne 0$ cannot cross the Hubble radius. However, for the $l=0$ modes, we get $V_{l}(R,H) = 0$ altogether, and hence these $s-$wave modes see no potential whatsoever, and they can simply pass from the sub-Hubble to the super-Hubble region or vice versa. For this study, we shall only consider the $l=0$ modes, i.e., the $s-$ waves to discuss the physics associated with particle creation. For the $s-$wave approximation, we have
\begin{equation}
    \Bigg(\frac{\partial^2\Phi^{00}_\Omega}{\partial T_*^2}- \frac{\partial^2\Phi^{00}_\Omega}{\partial R^2}\Bigg) = 0.
\end{equation}
The field expansion is written as
\begin{equation}\label{FieldExpansionTRFrame}
    \Phi = \int d^3\Omega (U^{\text{sub}}_\Omega b^L_{\Omega>\Omega_H}+V^{\text{sub}}_\Omega b^R_{\Omega>\Omega_H}) + \text{h.c.}
\end{equation}
The mode functions are 
\begin{equation}\label{USubMode}
    U_\Omega = \frac{1}{4\pi\sqrt{\Omega}R}e^{-i\Omega(T_*-R)}
\end{equation}
and 
\begin{equation}\label{VSubMode}
    V_\Omega = \frac{1}{4\pi\sqrt{\Omega}R}e^{-i\Omega(T_*+R)}
\end{equation}
and they are valid for all points including $R= 1/H$. We shall use these modes to calculate the Bogolyubov coefficient and power spectrum. Finally, we need to transfer the frequencies ($\Omega$) defined with respect to time  $(T_*)$ to the de Sitter static time $T$.

Before we make progress to calculate the beta Bogolyubov coefficient, let us come back to our methodology regarding using above mode functions. Consider the positive frequency mode given by $e^{-i\Omega T_*}$ with respect to the time $T_*$. If we operate the Hamiltonian operator $i\hbar \left(\partial/{\partial T}\right)$ on the aforementioned mode function, we obtain
\ber
i\hbar \frac{\partial}{\partial T} e^{-i\Omega T_*} &=& 2\hbar e^{2HT+1}\Omega e^{-i\Omega T_*}\\
                                                     &=& \omega(T) e^{-i\Omega T_*} \label{btrans},
\eer
where the time-dependent frequency $\omega(T)=2e^{2HT+1}\Omega$. Therefore, for $\Omega>0$, and for any given time $T$, we also have $\omega(T)>0$. That is we can also consider the mode  $e^{-i\Omega T_*}$ as positive frequency but with an instantaneous time-dependent frequency. Therefore, it is possible to calculate the Bogolyubov coefficient using the modes $e^{-i\Omega T_*}$, but finally we need to recast everything by changing $\Omega\rightarrow \omega(T)$, so that the final result holds for the static de Sitter frame.


\subsection{Bogolyubov coefficients}
We now know the field modes in the comoving and \textit{static de Sitter} frames. Consider the ingoing modes in the sub-Hubble region; these are given by
\begin{equation}
    f_k^{(u)\text{R}}=\frac{i}{(2k)^{3/2}}\left[C_k \frac{e^{iku}}{R} - D_k \frac{e^{-iku}}{R}\right]
\end{equation}
in the comoving frame and
\begin{equation}
    U_\Omega = \frac{1}{\sqrt{2\Omega}}e^{-i\Omega U_*}
\end{equation}
in the \textit{static de Sitter} frame, where $U_*=T_*-R$. Then the relevant Bogoliubov coefficient
\be
\beta_{\Omega k} = -i\int dU_*(f_k^{(u)\text{R}}\partial_{U_*}U_\Omega-U_\Omega\partial_{U_*}f_k^{(u)\text{R}})
\ee
where $U_* = T_*-R$ and it is easy to find the relationship of $U_*$ with the cosmological null coordinate $u=\eta - r$ as
\be
U = \pm \frac{{\cal H}e}{2} u^{2}.
\ee
Then we can rewrite the above integral as
\ber
\beta_{\Omega k}&=&-iAC_k\left[\int_0^\infty du\left(\left(\Omega u +\f{k}{\cH e}\right)e^{i(ku-Bu^2)}\right)+H.C.\right]\\
&&+iAD_k\left[\int_0^\infty du\left(\left(\Omega u -\f{k}{\cH e}\right)e^{-i(ku+Bu^2)}\right)+H.C.\right]
\eer
This leads to the following expression
\ber
\beta_{\Omega k}&=&\left[\f{1}{\cH e}(2C_k+D_k)\right]\sqrt{\f{\cH e}{2k}}\f{1}{2\Omega}\sin\left(\f{k^2}{2\cH e\Omega}-\f\pi4\right)\Gamma\left[\f12,\f{k^2}{2\cH e\Omega}\right] \nonumber\\
&&-\left[\f{D_k}{\cH e}+\f1k\sqrt{\f{2\Omega}{\cH e}}(C_k-D_k)\right]\sqrt{\f{\cH e}{2k}}\f{1}{2\Omega}\cos\left(\f{k^2}{2\cH e\Omega}\right)\Gamma\left[1,\f{k^2}{2\cH e\Omega}\right].\nonumber\\
\eer

Once again, considering the two choices of the vacuum states, one has an option to choose the coefficients $C_k$ and $D_k$ for the Bunch-Davies or the comoving vacuum states and obtain corresponding expressions for the Bogolyubov coefficients. First, let us consider the Bunch-Davies values of $C_k$ and $D_k$, and then, the modulus square gives
\ber
|\beta_{\Omega k}|^2&=&\left[\f{1}{4k^3\sqrt{2k\cH}\Omega}e^{-\f{k^2}{2e\cH\Omega}}\left[\f{k}{e}(e\cH^2-2k^2)\cos\left[\f{2k}{\cH}\left(1+\f{2}{\sqrt{e}}\right)\right]\right.\right.\nonumber\\
&&+\f{2\cH k^2}{\sqrt{e}}\sin\left[\f{2k}{\cH}\left(1+\f{2}{\sqrt{e}}\right)\right]-\sqrt{2\cH\Omega}e\cH^2\cos\f{k}{\cH}\nonumber\\
&&-\sqrt{2\cH\Omega}(e\cH^2-2k^2)\cos\left[\f{k}{\cH}\left(1-\f{2}{\sqrt{e}}\right)\right]\nonumber\\
&&\left.+2k\cH\sqrt{2e\cH\Omega}\sin\left[\f{k}{\cH}\left(1-\f{2}{\sqrt{e}}\right]\right)\right]\cos\f{k^2}{2e\cH\Omega}\nonumber\\
&&+\left[-\f{ke\cH^2}{2k^3\sqrt{2e\cH k}\Omega}\cos\f{k}{\cH}+\f{1}{4k^2\sqrt{2e\cH k}\Omega}(e\cH^2-2k^2)\cos\left[\f{k}{\cH}\left(1-\f{2}{\sqrt{e}}\right)\right]\right.\nonumber\\
&&\left.\left.-\f{k^2\cH}{2k^3\sqrt{2k\cH}\Omega}\sin\left[\f{k}{\cH}\left(1-\f{2}{\sqrt{e}}\right)\right]\right]\cos\left(\f{k^2}{2e\cH\Omega}+\f\pi4\right)\Gamma\left[\f12,\f{k^2}{2e\cH\Omega}\right]\right]^2\nonumber\\
&&+\left[\f{1}{4k^3\sqrt{2k\cH}\Omega}e^{-\f{k^2}{2e\cH\Omega}}\left[\f{k}{e}(e\cH^2-2k^2)\sin\left[\f{2k}{\cH}\left(1+\f{2}{\sqrt{e}}\right)\right]\right.\right.\nonumber\\
&&-\f{2\cH k^2}{\sqrt{e}}\cos\left[\f{2k}{\cH}\left(1+\f{2}{\sqrt{e}}\right)\right]+\sqrt{2\cH\Omega}e\cH^2\cos\f{k}{\cH}\nonumber\\
&&+\sqrt{2\cH\Omega}(e\cH^2-2k^2)\sin\left[\f{k}{\cH}\left(1-\f{2}{\sqrt{e}}\right)\right]\nonumber\\
&&\left.+2k\cH\sqrt{2e\cH\Omega}\cos\left[\f{k}{\cH}\left(1-\f{2}{\sqrt{e}}\right]\right)\right]\cos\f{k^2}{2e\cH\Omega}\nonumber\\
&&+\left[\f{ke\cH^2}{2k^3\sqrt{2e\cH k}\Omega}\sin\f{k}{\cH}-\f{1}{4k^2\sqrt{2e\cH k}\Omega}(e\cH^2-2k^2)\sin\left[\f{k}{\cH}\left(1-\f{2}{\sqrt{e}}\right)\right]\right.\nonumber\\
&&\left.\left.-\f{k^2\cH}{2k^3\sqrt{2k\cH}\Omega}\cos\left[\f{k}{\cH}\left(1-\f{2}{\sqrt{e}}\right)\right]\right]\cos\left(\f{k^2}{2e\cH\Omega}+\f\pi4\right)\Gamma\left[\f12,\f{k^2}{2e\cH\Omega}\right]\right]^2.\nonumber\\
\eer
The physical Bogolyubov coefficient valid for the static de Sitter observer living the radiation stage can now be found simply by replacing $\Omega\rightarrow \omega(T)$ using \eqref{btrans} which would provide us an instantaneous expression for $|\beta_{\omega k}|^2$ and hence an instantaneous particle number density.

\section{Summary and Conclusions}
\label{con}
To summarize, we have extended some physical aspects of the quantum field theory, usually studied in the inflationary de Sitter epoch, to the next expansion stage, given by the radiation domination. Particularly, our aim was to calculate the power spectrum resulted from the initial conditions on the state of the quantum field set in the inflationary universe. We succeeded to derive an expression for the power spectrum, given by \eqref{psrd00}, which is defined in the radiation stage and can be expressed for both choices of initial state of the field (i.e., Bunch-Davies or comoving vacuum). We also depicted this power spectrum and compared them for these two choices of the vacuum states. Another new result was presented in section \ref{dsinrad} where the static de Sitter frame was extended to the radiation stage which is described as a non-static spacetime and given by \eqref{mt4}. Finally, we used a conformally static form of the radiation dominated universe \eqref{statric},  which is just a time rescaling of \eqref{mt4}, for the calculation of the Bogolyubov coefficient representing the particle creation. In fact, we get an insight of time-dependent nature of particle number density in the radiation stage because of this study. In all cases, we see that the thermal or near thermal behavior in the inflationary de Sitter stage gets destroyed by the transition to the radiation stage. This was quite expected, but what was not known was how these expressions would turn out to be, and we now show exactly how these modification takes place. We believe this might be helpful to understand the physics of early universe in an unified manner where individual epochs and their transitions have physical implications. In future, we would like to extend our study to the subsequent stages of expansions even beyond the radiation stage.

\section{Acknowledgements}
The research of JRS and SKM are supported by the CONACyT Project CB/2017-18/A1S-33440. SKM also acknowledge the support of CONACyT ``Ciencia Frontera'' project 140630. The authors dedicate this work to the fond memory of Professor Thanu Padmanabhan who was a collaborator on this project for a brief amount of time.

\end{document}